\documentclass[twocolumn]{aastex62}

\usepackage{graphics,epsf}
\usepackage{amsmath}                % American Mathematical Society package
\usepackage{amsfonts}               % American Mathematical Society fonts
\usepackage{amssymb}                % American Mathematical Society symbol
\usepackage{epsfig}                 % EPS figures
\usepackage{graphicx}
\usepackage{float}
\usepackage{color}
\usepackage[para,online,flushleft]{threeparttable}
\def \cm{~\rm{cm}}
\def \s{~\rm{s}}
\def \km{~\rm{km}}

\def \g{~\rm{g}}

\def \AU{~\rm{AU}}

\def \yr{~\rm{yr}}

\begin{document}

\title{A pre-explosion extended effervescent zone around core collapse supernova progenitors} 

%% \correspondingauthor{Noam Soker}
%% \email{soker@physics.technion.ac.il}

%% \author{Efrat Sabach}
%%% \affiliation{Department of Physics, Technion, Haifa, 3200003, Israel}

\author[0000-0003-0375-8987]{Noam Soker}
\affiliation{Department of Physics, Technion, Haifa, 3200003, Israel; soker@physics.technion.ac.il}
\affiliation{Guangdong Technion Israel Institute of Technology, Shantou 515069, Guangdong Province, China}

\begin{abstract}
I propose a scenario according to which the dense compact circumstellar matter (CSM) that the ejecta of many core collapse supernovae (CCSNe) collide with within several days after explosion results from a dense zone where in addition to the stellar wind there is gas that does not reach the escape velocity. In this effervescent zone around red supergiant (RSG) stars, there are dense clumps that are ejected from the vicinity of the RSG surface, rise to radii of tens of astronomical units, and then fall back. I consider two simple velocity distributions of the ejected clumps. I find that the density of the bound mass can be tens of times that of the escaping wind, and therefore can mimic a very high mass loss rate.    
The dense effervescent compact CSM zone can (1) explain the collision of the ejecta of many CCSNe with a dense compact CSM days after explosion, (2) facilitate very high mass loss rate if the star experiences powerful pre-explosion activity, (3) form dust that obscures the progenitor in the visible band, and (4) lead to an efficient mass transfer to a stellar companion at separations of tens of astronomical units, if exists. The effervescent zone might exist for thousands of years and more, and therefore the effervescent CSM model removes the requirement from many type II CCSN progenitors to experience a very strong outburst just years to months before explosion.   
\end{abstract}


% ==========================================================
\section{Introduction}
\label{sec:intro}
% ==========================================================

There are two types of indications to the presence of compact circumstellar matter (CSM) around many, but not all, progenitors of core collapse supernovae (CCSNe) at explosion. One indication comes from pre-explosion outbursts years to months before the explosion and the other is the collision of the CCSN ejecta with compact CSM.

CCSNe are the explosions of stars that leave behind either a neutron star or a black hole. Stars in the initial mass range of $\ga 10 M_\odot$ form an iron core at the end of their nuclear burning phases. When the mass of the iron core grows to about $1.2 M_\odot$ the core collapses to form a NS (or with more mass a black hole). A small fraction of the gravitational energy that the formation of the neutron star releases explodes the rest of the star. Stars in the initial mass range of $\simeq 8M_\odot$ to $\simeq 10 M_\odot$ might collapse at an earlier phase, when their core is made up from ONeMg. Electron capture by magnesium in the degenerate core reduces the pressure and leads to core collapse, leaving behind a neutron star. 
In this study I consider the compact CSM around red supergiant (RSG) progenitors of both types of core collapse. 

Pre-explosion outbursts that are accompanied by high mass loss rate episodes might occur tens of years to only days prior to explosion (e.g., \citealt{Foleyetal2007, Pastorelloetal2007, Smithetal2010, Pastorelloetal2013, Marguttietal2014, Ofeketal2014, SvirskiNakar2014, Tartagliaetal2016, Strotjohannetal2021}). In some cases the pre-explosion outburst might take place as early as carbon-burning phase (e.g., \citealt{Moriyaetal2014, Marguttietal2017}).

\cite{Yaronetal2017} argue that the progenitor of SN~2013fs had an enhanced mass loss rate of $\simeq 0.3-4 \times 10^{-3} M_\odot \yr^{-1}$ for a wind velocity of $v_{\rm w} \approx 15-100 \km \s^{-1}$. They find the wind velocity to be $v_{\rm w} \la 100 \km \s^{-1}$.
\cite{Hosseinzadehetal2018} estimate the mass of the compact CSM around the progenitor of SN~2016bkv to be $\approx 0.04 M_\odot$. 
The typical sizes of the compact CSM that, e.g., \cite{Bruchetal2020}, find around CCSNe, on the other hand, are $R_{\rm CSM} \approx 10^{15} \cm$. If this compact CSM is an outflowing wind, the high mass loss rate begins at about $\simeq R_{\rm CSM}/v_{\rm w} \approx 3 \yr$, where $v_{\rm w} \simeq 100 \km \s^{-1}$ is the  wind velocity. In some cases the pre-explosion wind starts only several months before explosion (e.g., \citealt{Bruchetal2020}). \cite{Prenticeetal2020} claim that the CSM around the envelope-stripped SN~2018gjx was ejected within four months from explosion. In that case the CSM is non-spherical, most likely indicating a pre-explosion binary interaction \citep{Prenticeetal2020}.  
 
There are theoretical studies that include an enhanced mass loss rate years to months before explosion as a result of core activity (e.g., \citealt{Morozovaetal2020}), either set by excitation of waves (e.g., \citealt{QuataertShiode2012, ShiodeQuataert2014, Fuller2017, FullerRo2018}) or by core-magnetic activity (e.g., \citealt{SokerGilkis2017}). 
There are two problems with such a core activity as an explanation to all cases of SNe II with compact CSM. 
(1) In many cases this core activity might lead to substantial envelope expansion, but without much mass loss rate enhancement (e.g., \citealt{McleySoker2014}). 
(2) \cite{Bruchetal2020} find that $>30\%$ of SNe~II have compact CSM, while the fraction of CCSNe that suffer a pre-explosion outburst, which the core activity should excite, is only $\approx 10 \%$ (e.g., \citealt{Marguttietal2017}).

Although pre-explosion outbursts do occur, in the present study I consider another model to account for the frequent presence of compact CSM, the \textit{effervescent CSM model}. In this model there is a dense bound mass in an extended zone around the RSG surface (an extended zone relative to the stellar radius, but at the same time a compact CSM). \cite{Dessartetal2017} presented the idea of a long-lived complex extended dense zone around RSG progenitors of CCSNe that might have up to $\approx 0.01 M_\odot$. There are some similarities between their model and the effervescent CSM model.  
Before presenting the effervescent CSM model for CCSN progenitors, I turn to describe the motivation for introducing this model. 
     
In low-mass asymptotic giant branch (AGB) stars that are potential progenitors of planetary nebulae there are several observations that have lead to the development of the effervescent CSM (wind) model \citep{Soker2008eff}. These observations include complicated structures of SiO maser clumps (e.g., \citealt{Cottonetal2006}) and a chaotic inflow-outflow motion around the surface of some AGB stars (e.g., \citealt{DiamondKemball2003}).
Water maser observations that explore regions at larger distances from the surface of some AGB stars, $r \approx 100 \AU$, also indicate inhomogeneous outflows (e.g., \citealt{Vlemmingsetal2002}). The distribution of dust close to some AGB stars is also inhomogeneous, and might be related to the magnetic field in the atmosphere of AGB stars (e.g., \citealt{Khourietal2020} and references therein). Mira~A is a pulsating AGB star with a radius of $R_1 \simeq 500 R_\odot$ (e.g., \citealt{WoodKarovska2006}), and with an inhomogeneous and clumpy asymmetrical tens-AUs extended zone, that is, a compact CSM (e.g., \citealt{Planesasetal1990, RydeSchoier2001, Lopezetal1997}). \cite{Lopezetal1997} considered a model where dusty clumps $\approx 100$ times denser than their environment exist at tens of AUs from the AGB star Mira~A. They assume these clumps to explain the  IR emission, but did not consider the motion of the clumps. 
   
The carbon AGB star IRC+10216 further motivates the introduction of the effervescent CSM model as it has clumps within $\approx 10 \AU$ that \cite{Fonfriaetal2008} suggest move outward and inward at high velocities along different radial directions; the fast outward moving clumps reaches distances of $\gg 10 \AU$.  The post-AGB star HD56126 has both outflowing and inflowing gas around it at velocities of up to $\simeq 20 \km \s^{-1}$ \citep{KlochkovaChentsov2008}. These velocities are non-negligible with respect to the escape velocity of $\simeq 60 \km \s^{-1}$ from this star \citep{Li2003}. 

More relevant to the present study are observations of inhomogeneous winds in RSGs (e.g., \citealt{LobelDupree2000, Humphreysetal2007}),  including massive dust clumps around the RSG VY~CMa \citep{Kaminski2019}. Observations (e.g., \citealt{JosselinPlez2007}) and theoretical studies (e.g., \citealt{Freytagetal2002}) suggest that such inhomogeneous winds might result from convective cells in the envelope and/or magnetic activity of the giant stars. 
\cite{BoianGroh2020} argue that CCSNe with CSM tend to come from high mass RSGs, i.e., having zero age main sequence mass of $M_{\rm ZAMS} \ga 15 M_\odot$. This also suggests luminous CCSN progenitors that can lift gas above the photosphere. 
There are studies of extended region where inflow and outflow coexist around stars that are close to their Eddington luminosity limit (e.g., \citealt{OwockivanMarle2008, vanMarleetal2009}). 

The above studies motivated me to introduce the effervescent CSM model to upper AGB stars \citep{Soker2008eff}. The basic postulate of this model is that the zone up to about 100 AUs is inhomogeneous and contains many clumps that do not escape the star, but rather fall back. 

\cite{Moriyaetal2017} consider a dense compact CSM that results from the acceleration zone of the wind. Namely, instead of a constant wind velocity they consider a wind with a velocity that increases with radius, a profile that makes the density higher than of a constant-velocity wind very close to the stellar surface (also \citealt{Moriyaetal2018}). The effervescent CSM model is different in some key aspects that I discuss later. 

\cite{Dessartetal2017} simulate CCSN explosion inside an extended complex CSM that is bound or a dense wind. Numerically, they took either an atmosphere with an extended scale height and/or a dense wind. The effervescent CSM model share some properties and implications with their complex extended zone.   
They cite as motivation for considering such an extended zone the  observations of the RSG Betelgeuse that show outflows and inflows (down-flows) out to several RSG stellar radii (e.g., \citealt{Ohnakaetal2011, Kervellaetal2016}). \cite{Dessartetal2017} noted that the extended zone removes the requirement for a fine-tuned stellar activity years to months before explosion. I also consider an extended zone, and in that I overlap with their study, but I consider the effervescent model of clumps that move out and fall back.
I borrow the effervescent CSM model from low mass AGB progenitors of planetary nebulae.    
Nonetheless, I actually strengthen the claim of \cite{Dessartetal2017} that an extended zone around the photosphere of many RSG progenitors of CCSNe might explain observations.
   
In section \ref{sec:Conditions} I present the basic condition for the presence of an effervescent CSM, and in section \ref{sec:SingleClump} I estimate the outer boundary of the effervescent zone. In section \ref{sec:UniformClumps} I present one type of model for the effervescent zone and in section \ref{sec:DensityOld} I present another type. I summarise in section \ref{sec:summary}. 

% =====================================================
\section{The conditions for a dense effervescent zone}
\label{sec:Conditions}
% =====================================================

Stellar pulsations, magnetic activity, rotation, and/or strong convection bring gas to the zone above and close to the photosphere. The stellar radiation cannot bring all this gas to escape, but brings a large fraction to almost escape velocities. This implies that the effervescent zone will be dense when the stellar radiation cannot accelerate most of the gas around the photosphere to escape velocity. In other words, the effervescent zone will become extended when the radiation momentum flux is about equal to the wind momentum flux. 
This might be the case during regular evolution of massive RSG stars, and more so for all RSG when powerful convection in the core (e.g., \citealt{QuataertShiode2012, Fuller2017, FullerRo2018}) and/or a powerful dynamo in the core (e.g., \citealt{SokerGilkis2017}) drive stronger convection in the envelope. In turn, the stronger envelope convection pushes more gas above the photosphere. This process does not require the power of these activities to be more than what convection in the envelope can carry as the direct mass ejection mechanism requires. For that, this process of mass lifting can start when core activity is still weak, implying many years, even thousands of years and more, before the explosion. There  is no need for a fine tuning of the activity to form the compact CSM.

At that stage when the radiation cannot accelerate most of the gas to the escape speed the mass loss rate in the wind is $\dot M_{\rm wc} \simeq \eta_{\rm w} L/(c v_w)$, where $v_w$ is the terminal wind speed, $L$ the stellar luminosity, and $\eta_{\rm w}$ is the average net effective times that a photon transfers momentum to the wind (only in the outward radial direction). In most cases $\eta_{\rm w} < 1$, but in dense and opaque winds it can be somewhat larger than $1$.  Substituting typical values gives
\begin{eqnarray}
\begin{aligned} 
\dot M_{\rm wc} & \simeq  4 \times 10^{-5} \eta_{\rm w}
\left( \frac{L}{2 \times 10^5 L_\odot} \right)
\\ & \times 
\left( \frac{v_w}{100 \km \s^{-1}} \right)^{-1}
M_\odot \yr^{-1} .
\label{eq:mwc1}
\end{aligned}
\end{eqnarray}
This shows that the effervescent zone becomes significant when mass loss rate is high. But when we consider the effervescent zone, the mass loss rate into the wind need not be as high as estimates that do not consider the effervescent zone require. 

% =====================================================
\section{A single clump}
\label{sec:SingleClump}
% =====================================================

\begin{figure}  [t]
\centering
\begin{tabular}{cc}
% [trim=left bottom right top, clip]
\includegraphics[trim=1.5cm 11.0cm 0cm 2.0cm ,clip, scale=0.45]{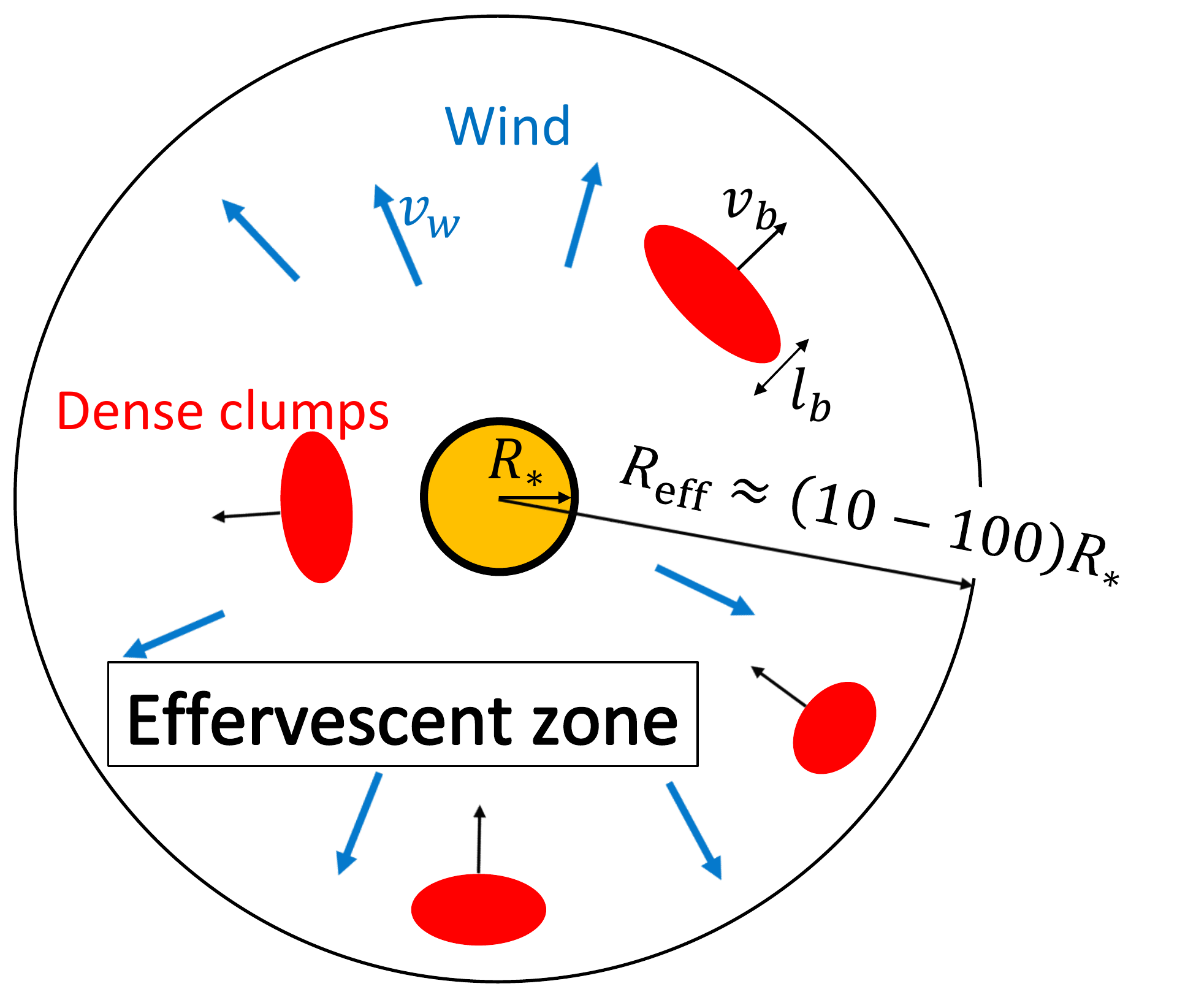} \\
\end{tabular}
% \vspace*{-2.5cm}
%\hskip -0.5 cm
\caption{A schematic drawing of the effervescent zone (not to scale). The thick-blue arrows depict the escaping wind at its more than escape velocity $v_w$. The red-oval clouds depict  the dense clumps that rise and fall within the effervescent zone. The orange sphere at the center is the RSG star of radius $R_\ast$. The outer edge of the effervescent zone is at $R_{\rm eff}$. 
}
  \label{fig:Effervescent}
    \end{figure}

Let us compare the different forces that act on a clump that is $k_b$ times denser than the ambient wind density, $\rho_b= k_b \rho_w$, and it moves at radial distances of tens of $\AU$, i.e., $r \gg R_\ast$, where $R_\ast$ is the stellar radius. The wind density is $\rho_w=\dot M_w/4 \pi r^2 v_w$, where $\dot M_w$ is the mass loss rate into the wind, and $v_w$ the wind velocity. 
I further characterise the clump with its cross section facing the star (perpendicular to the radial direction) $A_b$ and its length in the radial direction  $l_b$. I schematically draw the effervescent zone in Figure \ref{fig:Effervescent}.   
% FFFFFFFFFFFFFFFFFFFFFFFFFFFFFFFFFFFFFFFFFFFFFFFFFF
% FFFFFFFFFFFFFFFFFFFFFFFFFFFFFFFFFFFFFFFFFFFFFFFFFF 

The (radial) gravitational force on the clump due to the RSG star of mass $M_\ast$ is
\begin{equation}
F_g = - \frac{G M_\ast}{r^2}
\frac {\dot M_w}{4 \pi r^2 v_w}  k_b A_b l_b .
\label{eq:fgrav}
\end{equation}
The regular wind exerts drag force on the clump. In this simple treatment I take this force to be 
\begin{equation}
F_w \simeq \frac {\dot M_w v_w}{4 \pi r^2} A_b .
\label{eq:fwind}
\end{equation}
Namely, due to non-smooth clump surface I assume that the clump absorbs all the momentum of the wind that hits it. 
For an optical depth of $\tau_b$ along the radial direction of the clump the radiation exerts a force of 
\begin{equation}
F_{\rm rad} = A_b \frac{L}{4 \pi r^2 c} \left( 1-e^{-\tau_b} \right).
\label{eq:frad}
\end{equation}

I assumed above that the radiation pressure can accelerate both the wind that hits the clump, and the clump. The justification is that the radiation accelerates the wind within several stellar radii, whereas I consider the clumps at $r \ga 10 R_\ast$, where $R_\ast$ is the stellar radius. These distances are outside the winds' acceleration zone and the stellar radiation is close to the photospheric radiation. Namely, I assume that the wind is optically thin while the denser clumps are cooler and have dust. The opacity of the dusty clump is $\kappa_b \approx 10 \cm^2 \g^{-1}$, while that of the cool and partially neutral wind that contains much less dust is much lower. Overall, at tens of AUs the optical depth in the clumps is much larger than in the wind. However, further out the inner clumps make the radiation redder, and for this longer wavelength band the dust opacity becomes low. 
 
The condition on the blob acceleration to be down (in the $-r$ direction) is $F_g +  F_{\rm rad} + F_w <0$. Using equations (\ref{eq:fgrav})-(\ref{eq:frad}) I find this condition to read 
\begin{eqnarray}
\begin{aligned} 
1 > \frac{F_{\rm rad} + F_w }{-F_g}
& \simeq 
\frac{v^2_w}{v^2_{\rm Kep}} \frac {r}{R_\ast} \frac{r}{l_b} k^{-1}_b
\\ & \times 
\left[ \frac{L/c}{\dot M_w v_w}\left( 1-e^{-\tau_b} \right) + 1 \right],
\label{eq:Condition1}
\end{aligned} 
\end{eqnarray}
where $v_{\rm Kep}=\sqrt{G M_\ast/R_\ast}$ is the Keplerian velocity on the surface of the star. 

I assume that the clump expands radially, e.g., $A_b \propto r^2$, but that its radial length $l_b$ stays constant. I consider two limits. 
If the reddening of the stellar radiation by dust close to the star is significant, then the low dust opacity to this band implies that $\tau_b$ might be low. In addition, the low clump's density at large distances also reduces $\tau_b$. In that limit I neglect the first term in the square parenthesis of equation (\ref{eq:Condition1}), and so the requirement on the clump's acceleration to be negative (down to the star) reads  
\begin{equation}
\frac{r}{R_\ast} \la \frac{v_{\rm Kep}}{v_w} 
\left( \frac{l_b}{R_\ast} \right)^{1/2}    k^{1/2}_b.
\label{eq:Condition2}
\end{equation}
The demand on the clumps to form an extended effervescent zone is that the value of $k_b$ be large. 
I emphasise again here that it is not radiation pressure that eject the clumps from the stellar surface, but rather stellar pulsation, stellar convection, and magnetic activity in the outer envelope where gravity is relatively weak. 
 
If on the other hand radiation pressure exerts a larger outward force on the clump than that of the wind, then I approximate $( 1-e^{-\tau_b})^{1/2} \simeq 1$, and  the condition reads
 \begin{equation}
\frac{r}{R_\ast} \la \frac{v_{\rm Kep}}{v_w} 
\left( \frac{\dot M_w v_w}{L/c} \right)^{1/2}
\left( \frac{l_b}{R_\ast} \right)^{1/2}    k^{1/2}_b.
\label{eq:Condition3}
\end{equation}
Since we take $\dot M_w v_w \simeq L/c$ (equation \ref{eq:mwc1}), equations (\ref{eq:Condition2}) and (\ref{eq:Condition3}) are practically the same for the very evolved massive stars I study here. 

The conclusion is that to form an extended effervescent zone around the star the clumps should be hundreds to thousands of times denser than the average density of the wind. The extension of the effervescent zone is
\begin{eqnarray}
\begin{aligned}
R_{\rm eff} & \approx 6.7 \times 10^{14} \left( \frac{k_b}{1000} \right)^{1/2} 
\left( \frac{R_\ast}{2 \AU} \right) 
\\ & \times 
\left( \frac{l_b}{R_\ast} \right)^{1/2}  
\left( \frac{v_w}{v_{\rm esc}}  \right)^{-1} \cm ,
\label{eq:Reffer}
\end{aligned}
\end{eqnarray}
where I assume that the wind is saturated in the sense that $\dot M_w v_w \simeq L/c$ and that the wind speed is about the escape speed from the stellar surface $v_{\rm esc} = 2^{1/2} v_{\rm Kep}$. 

The density in the clump is 
\begin{eqnarray}
\begin{aligned} 
\rho_b  \simeq  & 9 \times 10^{-11} % 8.96
\left( \frac{\dot M_{\rm wc}}{4 \times 10^{-5}} \right) 
\left( \frac{r}{ 1 \AU} \right)^{-2} 
\\ & \times 
\left( \frac{v_w}{100 \km \s^{-1}} \right)^{-1}
\left( \frac{k_b}{1000} \right) \g \cm^{-3}.
\label{eq:DEnClump}
\end{aligned}
\end{eqnarray}
This density is more than one order of magnitude smaller than the density at the photosphere of RSG stars, $\rho_p$. As examples, a stellar model of zero age main sequence star of $M_{\rm ZAMS}=15 M_\odot$ has a photospheric density of 
$\rho_p(1 \AU)= 5 \times 10^{-9} \g \cm^{-3}$ and $\rho_p(2 \AU)= 3 \times 10^{-9} \g \cm^{-3}$, at stellar radii of $R_\ast =1 \AU$ and $R_\ast=2 \AU$, respectively, along its evolution (I obtained these values from simulating stellar models with the open stellar evolution code \textsc{mesa}; \citealt{Paxtonetal2018}). 
A stellar model of $M_{\rm ZAMS}=30 M_\odot$ has a photospheric density of $\rho_p(2 \AU)= 4 \times 10^{-9} \g \cm^{-3}$ and $\rho_p(5 \AU)= 6 \times 10^{-10} \g \cm^{-3}$, at stellar radii of $R_\ast =2 \AU$ and $R_\ast=5 \AU$, respectively. 
This implies that the star might lift such clumps above the photosphere. 

% =====================================================
\section{An effervescent zone from a group of unique velocity clumps}
\label{sec:UniformClumps}
% =====================================================

% =====================================================
\subsection{The average density}
\label{subsec:HighDensity}
% =====================================================
 
Consider a case when the activity in the envelope of the RSG star (pulsation, convection, rotation, magnetic activity, powerful radiation) ejects bound mass, i.e., with less than the escape velocity, at a rate of 
\begin{equation}
\dot M_{\rm eff} = \beta \dot M_{w}.  
\label{eq:beta}
\end{equation}
Consider also that the main force on the clumps is gravity, as the other forces that we consider in section \ref{sec:Conditions} play a role at very large radii where we assume that the effervescent zone is already depleted. The time it takes a clump to reach a maximum radius $R_{\rm eff}$ and fall back to the star is 
\begin{equation}
t_{\rm eff} \simeq \frac {2 \pi}{2^{3/2}} 
\frac {R_\ast}{v_{\rm Kep}} 
\left( \frac{R_{\rm eff}}{R_\ast} \right)^{3/2} 
\label{eq:Teff}. 
\end{equation}
A wind element spends a time of  $t_w (R_{\rm eff}) \simeq R_{\rm eff}/v_w$ in the effervescent zone. 
The ratio of the average density of the bound mass to that of the wind (total mass divided by total volume) in the effervescent zone is therefore 
\begin{eqnarray}
\begin{aligned}
\frac {\bar \rho_{\rm eff}}{\bar \rho_w} & \simeq 
\frac{\pi}{\sqrt{2}} 
\left( \frac{R_{\rm eff}}{R_\ast} \right)^{1/2} 
\left( \frac{v_w}{v_{\rm Kep}} \right)\beta
\\ & =
17
\left( \frac{R_{\rm eff}}{30 R_\ast} \right)^{1/2} 
\left( \frac{v_w}{v_{\rm esc}} \right) \beta, 
\label{eq:AverageDensityRatio} 
\end{aligned}
\end{eqnarray}
where in the second equality I scaled the wind velocity with the escape velocity from the RSG star.

Of course, many clumps will reach much smaller radii and spend much less time in the effervescent zone. On the other hand, for active RSG stars I expect $\beta >1$. Namely, most of the gas that the envelope activity lifts around the photosphere does not reach the escape velocity. 
The clumps move at a lower velocity in the outer regions of the effervescent zone, so the density ratio in the outer regions of the effervescent zone is larger even.
For example, $55 \%$ of its round trip the clump spends in the outer 20 percent of the effervescent zone, $0.8 R_{\rm eff} $ to $R_{\rm eff}$. 
For the same parameters I used in scaling equation (\ref{eq:AverageDensityRatio}) the ratio of average densities in this outer region of the effervescent zone, $\left( {\bar \rho_{\rm eff}}/{\bar \rho_w}\right)_{0.8-1} \simeq 47$.
% The program is: FallingTimeClumps.xlsx

I emphasise two points. (1) This large density ratio exists despite that the mass ejection rate into the effervescent zone is about equal to that in the wind, $\beta \simeq 1$. (2) The derivation in this section assumes that the RSG activity brings the wind to the escape velocity, but a similar amount of mass to be close to, but below, the escape velocity. The process of acceleration takes place along several stellar radii, and therefore the derivation here are approximate. 
Nonetheless, equation (\ref{eq:AverageDensityRatio}) does give the general behavior of the effervescent zone. 

% =====================================================
\subsection{The density profile}
\label{subsec:density}
% =====================================================
 
I consider that all clumps are ejected to radius $R_{\rm eff}$. In that case the average density at each radius, but not within a distance of $l_b$ from $R_{\rm eff}$ is 
\begin{equation}
\rho_{\rm eff} (r) \simeq 2 \frac {\dot M_{\rm eff}}{4 \pi r^2 v_b} \qquad {\rm for} \qquad R_\ast < r < R_{\rm eff} - l_b, 
\label{eq:rhoEff1} 
\end{equation}
where the factor 2 comes from that each clump moves outward and fall back. 
As above, I neglect the forces on the clumps due to radiation and the wind, and so the velocity of the clump is 
\begin{equation}
v_b \simeq \sqrt{ \frac {2 G M_\ast}{R_{\rm eff}} }
\sqrt {\frac{R_{\rm eff}}{r} -1}. 
\label{eq:rhoEff2}  
\end{equation}
 With the aid of equations (\ref{eq:beta}) and (\ref{eq:rhoEff2}), equation (\ref{eq:rhoEff1}) becomes 
\begin{eqnarray}
\begin{aligned}
& \rho_{\rm eff} (r) \simeq 2 \frac {\beta \dot M_{w}}{4 \pi r^2 v_{\rm esc}} 
\left( \frac {R_{\rm eff}}{R_\ast} \right)^{1/2}
\left( \frac {R_{\rm eff}}{r} - 1 \right)^{-1/2} 
\\ & =  2 \beta
\rho_{w} (r) \left( \frac{v_w}{v_{\rm esc}} \right)
\left( \frac {R_{\rm eff}}{R_\ast} \right)^{1/2}
\left( \frac {R_{\rm eff}}{r} - 1 \right)^{-1/2}   
\\ & 
\qquad {\rm for} \qquad R_\ast < r < R_{\rm eff} - l_b. 
\label{eq:rhoEff3} 
\end{aligned}
\end{eqnarray}

I note the following properties of the above density profile. 
\newline
(1) For $R_{\rm eff} \gg R_\ast$, as is the case in the present study, close to the star, i.e. at $r \simeq R_\ast$ the density of the bound gas is $\rho_{\rm eff}  \simeq 2 \beta \rho_w$. 
\newline
(2) For the above simple density profile case, the minimum value of the density is at $r=0.75 R_{\rm eff}$. At that radius 
\begin{equation}
\left( \rho_{\rm eff} \right)_{\rm min}= 
\rho_{\rm eff}(0.75 R_{\rm eff}) 
\simeq 19 \beta \left( \frac{R_{\rm eff}}{30 R_\ast} \right)^{1/2} \rho_{w}. 
\label{eq:rhoEffmin} 
\end{equation} 
This ratio increases to, e.g., $33$ at $r=0.9 R_{\rm eff}$, keeping other parameters the same. 
\newline
(3) The density profile is quite flat in an extended region (equation \ref{eq:rhoEff3}). As three examples,   
\begin{equation}
  \setlength{\arraycolsep}{0pt}
\rho_{\rm eff} = \rho_{\rm eff}(0.75 R_{\rm eff}) \times 
\left\{ \begin{array}{ l r }  
    1.57 &{} \quad  {\rm at} \quad r = 0.95 R_{\rm eff} \\
    1.19 &{} \quad  {\rm at} \quad r = 0.55 R_{\rm eff} \\
    1.66 &{} \quad  {\rm at} \quad r = 0.40 R_{\rm eff}     .
  \end{array} \right.
  \label{eq:rhoEff4} 
\end{equation}

Overall, if we consider the increase in density near $R_{\rm eff}$ that gives the high average density in the outer regions of the effervescent zone (equation \ref{eq:AverageDensityRatio} and the discussion below it), we see that the ejection at a mass rate about equal to that of the wind might mimic a short-lived wind with a mass loss rate that is tens times larger than the real mass loss rate.  

% ====================
\section{An example of velocity distribution}
\label{sec:DensityOld}
% ====================

In the first paper on the effervescent zone that aimed at AGB stars  (\citealt{Soker2008eff}) I concentrated on a simple derivation of a density profile for a specific case (for more details see that paper). I assumed that the bound gas is ejected from a radius of $R_0 \approx {\rm few} \times R_\ast$ (for the motivation to take $R_0 \approx {\rm few} \times R_\ast$ for the extended dense zone see \citealt{Soker2008eff} and \citealt{Dessartetal2017}), and I took the mass ejection rate in the velocity interval $v$ to $v+ dv$ as 
\begin{equation}
d \dot M_e = f \dot M_w \left( \frac{v}{v_{\rm esc,0}} \right)^{q} \frac{dv}{v_{\rm esc,0}} \quad {\rm for} \quad 0<v<v_m,
\label{eq:dmdot1}
\end{equation}
where $v_{\rm esc,0}$ is the escape speed from $R_0$ and $q$ ($-k_v$ in \citealt{Soker2008eff}) is a constant of the model. The maximum velocity $v_m$ is for clumps that reach the outer boundary of the effervescent zone $R_{\rm eff}$, with the relation $v_m={v_{\rm esc,0}} \left( 1-R_0/R_{\rm eff} \right)^{1/2}$ .
The total rate of unbound mass that the star ejects to the effervescence zone is 
\begin{eqnarray}
\begin{aligned}
\dot M_{\rm eff} & =
\frac{f}{1+q} \left( \frac{v_m}{v_{\rm esc,0}} \right)^{1+q} \dot M_w
\\ &  = 
\frac{f}{1+q} \left( 1 - \frac {R_0}{R_{\rm eff}} \right)^{\frac{1+q}{2}} \dot M_w,
\label{eq:Meff2}
\end{aligned}
\end{eqnarray}
From this, $f=(1+q) \left( 1-R_0/R_{\rm eff} \right)^{-(1+q)/2} \beta$, where I defined $\beta$ in equation (\ref{eq:beta}).

As I expect the very luminous RSG stars to lift more gas closer to the escape speed I take $q \gg 1$ (in the first paper I took low values of $q \approx 1$, and even negative values). Under the very crude assumption that each dense clump spends all the time at the maximum radius of its up and down trajectory I derived (\citealt{Soker2008eff}) the ratio of the bound density to the escaping wind density as 
\begin{eqnarray}
\begin{aligned}
& \frac{\rho_{\rm eff}}{\rho_{w}} 
\approx (1+q) \beta \left( 1 - \frac {R_0}{R_{\rm eff}} \right)^{-\frac{1+q}{2}}
\frac{\pi}{2} 
\left( \frac{v_w}{v_{\rm esc}} \right) 
\left( \frac{R_0}{R_\ast} \right)
\\ & \times 
\left( \frac{r}{R_\ast} \right)^{-1/2}
\left( 1- \frac{R_0}{r} \right)^{\frac{q-1}{2}}
\quad {\rm for} \quad R_0 < r <R_{\rm eff}, 
\label{eq:rhoEffPaperI} 
\end{aligned}
\end{eqnarray}
where as before $v_{\rm esc}$ is the escape velocity from the stellar surface. I scale equation (\ref{eq:rhoEffPaperI})  to allow comparison to equations (\ref{eq:rhoEffmin}) and (\ref{eq:rhoEff4})  
\begin{eqnarray}
\begin{aligned}
& \frac{\rho_{\rm eff}}{\rho_{w}} 
\approx 9.3 \frac{(1+q) \beta}{10} 
\left( \frac{v_w}{v_{\rm esc}} \right) 
\left( \frac{R_0}{3R_\ast} \right)
\left( \frac{r}{20R_\ast} \right)^{-1/2}
\\ & \times
\left[ \left( 1 - \frac {R_0}{R_{\rm eff}} \right)^{-\frac{1+q}{2}}
\frac{1}{1.69} \right] 
\\ & \times 
\left[ \left( 1- \frac{R_0}{r} \right)^{\frac{q-1}{2}} \frac{1}{0.52} \right]
\quad {\rm for} \quad R_0 < r <R_{\rm eff}, 
\label{eq:rhoEffPaperIScaled} 
\end{aligned}
\end{eqnarray}
where I normalised terms for $q=9$, $r=(20/3)R_0=20 R_\ast$, and $R_{\rm eff}= 10 R_0 = 30 R_\ast$ (for $R_0=3 R_\ast$). 

Keeping all other parameters the same, the ratio in equation (\ref{eq:rhoEffPaperIScaled}) at two other radii are ${\rho_{\rm eff}}/{\rho_{w}} \approx 8.4$ and $9.6$, at $r= 5 R_0 = 15 R_\ast$ and $r=10 R_0=R_{\rm eff}$, respectively. Over all, the density in the outer regions of this version of the effervescent zone decreases more or less as $r^{-2}$, as the wind does.
To increase the density in the outer regions of the effervescent zone to tens times the wind density for this distribution of clump velocities, would require taking $\beta >1 $ and/or increasing further the value of $q$.  

% =====================================================
\section{Summary and implications}
\label{sec:summary}
% =====================================================

I presented a phenomenological model for an extended high density zone around RSG stars at the end of their life. The basic assumption is that dense clumps, much denser than the escaping wind, move up and fall back. I term this the effervescent CSM model. The effervescent zone extends to tens of stellar radii, i.e., to $R_{\rm eff} \approx 10-100 \AU \approx 10^{14} - 10^{15} \cm$ for RSG stars. Such an effervescent zone might exist only when the RSG is large, such that surface gravity is relatively weak, and the radial momentum flux of the (escaping) wind is about equal to the momentum flux of the stellar radiation (equation \ref{eq:mwc1}). 
 
The effervescent CSM model assumes that during such a stellar evolutionary phase the stellar activity (rotation, convection, magnetic activity, disturbances from the vigorous core nuclear burning; see section\ref{sec:intro}) lifts large amounts of mass above the photosphere, but the stellar radiation manages to unbound only about half of that mass or less. The rest almost escapes, but falls back after reaching large radii. To fall back rather than be accelerated by radiation and the wind, the effervescent zone extends to no more than tens to about one hundred of AUs (equation \ref{eq:Reffer}).    
  
I considered two phenomenological distributions of clumps' velocities (after accelerated out). In the first one (section \ref{sec:UniformClumps}), all clumps move at one velocity very close to the escape velocity. I found that the average density of the bound gas (equation \ref{eq:AverageDensityRatio}) is tens times larger than that of the escaping wind, in particular in the outer regions of the effervescent zone (equations \ref{eq:rhoEffmin}, \ref{eq:rhoEff4}). 
 
For example, consider a case where the outer radius of the effervescent zone is 20 times the stellar radius of a large RSG star of stellar radius $R_{\ast} =2 \AU$, i.e., $R_{\rm eff}\simeq 20 R_\ast \simeq 40 \AU \simeq 6 \times 10^{14} \cm$. For the escaping wind properties as given by equation (\ref{eq:mwc1}), which for $v_w=v_{\rm esc}$ corresponds to an RSG star of mass $M_\ast=11.3 M_\odot$, the total wind mass inside $r<R_{\rm eff}$ is $M_w(R_{\rm eff}) \simeq 7 \times 10^{-5} M_\odot$. However, the mass of the bound gas is $14 \beta$ times larger (equation \ref{eq:AverageDensityRatio}), i.e., $M_{\rm eff} \simeq 0.001 \beta M_\odot$. However, in the very outer parts of the effervescent zone this ratio can be as large as $\approx 100$, mimicking a mass loss rate in the last year or so of $\approx 0.001-0.01 M_\odot \yr^{-1}$.
This might explain some cases of CCSNe with compact CSM, e.g., 
in SN~2013fs for which \cite{Yaronetal2017} estimated an enhanced mass loss rate of $\simeq 0.3-4 \times 10^{-3} M_\odot \yr^{-1}$. 

\cite{Dessartetal2017} already considered a dense bound gas around RSG progenitors of CCSNe. In the present study I presented calculations of outflowing and in-flowing gas to obtain the density profile, rather than assuming it or taking an extended stellar atmosphere. 
\cite{Dessartetal2017} estimate the mass around the progenitor of SN~2013fs to have been $\approx 0.01 M_\odot$ spread over $\approx 2 \times 10^{14} \cm$. To account for this amount of mass we would require $\beta \approx 10$ in the frame of the effervescent CSM model that I studeid here. 
To account for a compact CSM of $0.04 M_\odot$ as in SN~2016bkv \citep{Hosseinzadehetal2018} or of $0.07 M_\odot$ as in SN~2018cuf \citep{Dongetal2021}, we would require to take $\beta \simeq 10$. It is a large value, but not unreasonable in the effervescent model if the RSG experiences strong stellar activity, e.g., due to rapid rotation. 

In section \ref{sec:DensityOld} I adopted the velocity distribution of the clumps (equation \ref{eq:dmdot1}) from my earlier paper \citep{Soker2008eff}. I followed the treatment from that paper, and derived the density profile for this clumps' velocities distribution in equations (\ref{eq:rhoEffPaperI}) and (\ref{eq:rhoEffPaperIScaled}). This case requires values of $\beta >1$ to achieve high densities in the outer effervescent zone.  
 
\cite{Prenticeetal2020} study the type IIb (envelope-stripped) SN~2018gjx and argue that its CSM was non-spherical, probably a torus, and extended to about $20-30 \AU$ at explosion. They further estimate the mass in the CSM to be $\approx 0.004-0.014 M_\odot$ and the mass loss rate from the progenitor that formed this CSM to be $\approx 0.01-0.05 M_\odot \yr^{-1}$. If indeed the CSM is in a torus, this geometry suggests a strong binary interaction (e.g., \citealt{GofmanSoker2019}) that probably spun-up the progenitor. With rapid rotation the stellar progenitor might be able to lift a dense equatorial effervescent zone with the required mass. I therefore raise the possibility that even in these non type II CCSN the compact CSM might be an effervescent zone rather than an intensive escaping wind. 

The main motivation to consider the effervescent zone in RSG stars is to remove both the requirement for fine-tuned and strong stellar activity years to months before explosion. The effervescent CSM model can exist for a long time, thousands of years and more before explosion, and the flow structure by which the gas does not reach the escape speed does not require strong activity. Moderate activity due to some extra rotation and/or moderate core activity might be sufficient. 

Another effect of the effervescent zone is an extra mass transfer to a companion, if exists within and close to the effervescent zone \citep{Soker2008eff}. To have a high mass transfer rate it is sufficient that the effervescent zone overfills the Roche lobe of the RSG star, such that mass transfer through the first Lagrangian point takes place
(\citealt{Harpazetal1997, MohamedPodsiadlowski2007, MohamedPodsiadlowski2012, Chenetal2017, Saladinoetal2018, Chenetal2020}; this is also termed wind-Roche lobe overflow). Due to the high specific angular momentum of the transferred mass, it will form an accretion disk around the secondary star (if it is not a giant). In turn, the accretion disk 
is likely to launch two jets that will shape the CSM (e.g., \citealt{Hilleletal2020}).

% ===================================================
\section*{Acknowledgments}
% ===================================================
 
I thank Avishai Gilkis, Amit Kashi and an anonymous referee for useful comments. I thank Aldana Grichener for simulating stellar models. This research was supported by a grant from the Israel Science Foundation (420/16 and 769/20) and a grant from the Asher Space Research Fund at the Technion.

% Calculation of falling clumps in main/zFAllingTimeClumps.xlsx
% =======================


\begin{thebibliography}


\bibitem[Boian \& Groh(2020)]{BoianGroh2020} Boian, I. \& Groh, J.~H.\ 2020, \mnras, 496, 1325
 
 \bibitem[Bruch et al.(2020)]{Bruchetal2020} Bruch, R.~J., Gal-Yam, A., Schulze, S., et al.\ 2020, arXiv:2008.09986

\bibitem[Chen et al.(2017)]{Chenetal2017}  Chen, Z., Frank, A., Blackman, E.~G., Nordhaus, J., \& Carroll-Nellenback J.\ 2017, \mnras, 468, 4465 

\bibitem[Chen et al.(2020)]{Chenetal2020} Chen, Z., Ivanova, N., \& Carroll-Nellenback, J.\ 2020, \apj, 892, 110

\bibitem[Cotton et al.(2006)]{Cottonetal2006} Cotton, W.~D., Vlemmings, W., Mennesson, B., et al.\ 2006, \aap, 456, 339

\bibitem[Dessart et al.(2017)]{Dessartetal2017} Dessart, L., John Hillier, D., \& Audit, E.\ 2017, \aap, 605, A83

\bibitem[Diamond \& Kemball(2003)]{DiamondKemball2003} Diamond, P.~J. \& Kemball, A.~J.\ 2003, \apj, 599, 1372

\bibitem[Dong et al.(2021)]{Dongetal2021}  Dong, Y., Valenti, S., Bostroem, K.~A., et al.\ 2021, arXiv:2010.09764 

\bibitem[Foley et al.(2007)]{Foleyetal2007} Foley, R.~J., Smith, N., Ganeshalingam, M., Li, W., Chornock, R., \& Filippenko, A.~V.\ 2007, \apjl, 657, L105


\bibitem[Fonfr{\'\i}a et al.(2008)]{Fonfriaetal2008} Fonfr{\'\i}a, J.~P., Cernicharo, J., Richter, M.~J., \& Lacy, J. H.\ 2008, \apj, 673, 445

\bibitem[Freytag et al.(2002)]{Freytagetal2002} Freytag, B., Steffen, M., \& Dorch, B.\ 2002, Astronomische Nachrichten, 323, 213

\bibitem[Fuller(2017)]{Fuller2017} Fuller, J.\ 2017, \mnras, 470, 1642
    
\bibitem[Fuller \& Ro(2018)]{FullerRo2018} Fuller, J. \& Ro, S.\ 2018, \mnras, 476, 1853

\bibitem[Gofman \& Soker(2019)]{GofmanSoker2019} Gofman, R.~A. \& Soker, N.\ 2019, \mnras, 488, 5854
 
\bibitem[Harpaz et al.(1997)]{Harpazetal1997} Harpaz, A., Rappaport, S., \& Soker, N.\ 1997, \apj, 487, 809 
 
\bibitem[Hillel et al.(2020)]{Hilleletal2020} Hillel, S., Schreier, R., \& Soker, N.\ 2020, \apj, 891, 33

\bibitem[Hosseinzadeh et al.(2018)]{Hosseinzadehetal2018} Hosseinzadeh, G., Valenti, S., McCully, C., et al.\ 2018, \apj, 861, 63

\bibitem[Humphreys et al.(2007)]{Humphreysetal2007} Humphreys, R.~M., Helton, L.~A., \& Jones, T.~J.\ 2007, \aj, 133, 2716

\bibitem[Josselin \& Plez(2007)]{JosselinPlez2007} Josselin, E. \& Plez, B.\ 2007, \aap, 469, 671

\bibitem[Kami{\'n}ski(2019)]{Kaminski2019} Kami{\'n}ski, T.\ 2019, \aap, 627, A114

\bibitem[Kervella et al.(2016)]{Kervellaetal2016} Kervella, P., Lagadec, E., Montarg{\`e}s, M., et al.\ 2016, \aap, 585, A28

\bibitem[Khouri et al.(2020)]{Khourietal2020} Khouri, T., Vlemmings, W.~H.~T., Paladini, C., et al.\ 2020, \aap, 635, A200

\bibitem[Klochkova \& Chentsov(2007)]{KlochkovaChentsov2008} Klochkova, V.~G. \& Chentsov, Y.~L.\ 2007, Astronomy Reports, 51, 994

\bibitem[Li(2003)]{Li2003} Li, A.\ 2003, \apjl, 599, L45
 
\bibitem[Lobel \& Dupree(2000)]{LobelDupree2000} Lobel, A. \& Dupree, A.~K.\ 2000, \apj, 545, 454

\bibitem[Lopez et al.(1997)]{Lopezetal1997} Lopez, B., Danchi, W.~C., Bester, M., et al.\ 1997, \apj, 488, 807

\bibitem[Margutti et al.(2017)]{Marguttietal2017} Margutti, R., Kamble, A., Milisavljevic, D., et al.\ 2017, \apj, 835, 140


\bibitem[Margutti et al.(2014)]{Marguttietal2014} Margutti, R., Milisavljevic, D., Soderberg, A.~M., et al.\ 2014, \apj, 780, 21

\bibitem[Mcley \& Soker(2014)]{McleySoker2014} Mcley, L., \& Soker, N.\ 2014, \mnras, 445, 2492

\bibitem[Mohamed \& Podsiadlowski(2007)]{MohamedPodsiadlowski2007} Mohamed, S., \& Podsiadlowski, P.\ 2007, 15th European Workshop on White Dwarfs, 372, 397 

\bibitem[Mohamed \& Podsiadlowski(2012)]{MohamedPodsiadlowski2012} Mohamed, S., \& Podsiadlowski, P.\ 2012, Baltic Astronomy, 21, 88

\bibitem[Moriya et al.(2018)]{Moriyaetal2018} Moriya, T.~J., F{\"o}rster, F., Yoon, S.-C., Gr{\"a}fener, G., \&  Blinnikov, S.~I.,\ 2018, \mnras, 476, 2840

\bibitem[Moriya et al.(2014)]{Moriyaetal2014} Moriya, T.~J., Maeda, K., Taddia, F., Sollerman, J., Blinnikov, S. I., Sorokina, E. I.\ 2014, \mnras, 439, 291

\bibitem[Moriya et al.(2017)]{Moriyaetal2017} Moriya, T.~J., Yoon, S.-C., Gr{\"a}fener, G., \& Blinnikov, S.~I.\ 2017, \mnras, 469, L108

\bibitem[Morozova et al.(2020)]{Morozovaetal2020} Morozova, V., Piro, A.~L., Fuller, J., \& Van Dyk S.~D.\ 2020, \apjl, 891, L32

\bibitem[Ofek et al.(2014)]{Ofeketal2014} Ofek, E.~O., Sullivan, M., Shaviv, N.~J., et al.\ 2014, \apj, 789, 104


\bibitem[Ohnaka et al.(2011)]{Ohnakaetal2011} Ohnaka, K., Weigelt, G., Millour, F., et al.\ 2011, \aap, 529, A163

\bibitem[Owocki \& van Marle(2008)]{OwockivanMarle2008} Owocki, S. P., \& van Marle, A. J., 2008, in Massive Stars as Cosmic Engines, IAU Symp 250, ed. F. Bresolin, P. A. Crowther, \& J. Puls (Cambridge Univ. Press) Volume 250, p. 71-82. 
       
\bibitem[Pastorello et al.(2013)]{Pastorelloetal2013} Pastorello, A., Cappellaro, E., Inserra, C., et al.\ 2013, \apj, 767, 1

\bibitem[Pastorello et al.(2007)]{Pastorelloetal2007} Pastorello, A., Smartt, S.~J., Mattila, S., et al.\ 2007, \nat, 447, 829

\bibitem[Paxton et al.(2018)]{Paxtonetal2018} Paxton, B., Schwab, J., Bauer, E.~B., et al.\ 2018, The Astrophysical Journal Supplement Series, 234, 34.

\bibitem[Planesas et al.(1990)]{Planesasetal1990} Planesas, P., Bachiller, R., Martin-Pintado, J., \& Bujarrabal, V.\ 1990, \apj, 351, 263

\bibitem[Prentice et al.(2020)]{Prenticeetal2020} Prentice, S.~J., Maguire, K., Boian, I., et al.\ 2020, arXiv:2009.10509

\bibitem[Quataert \& Shiode(2012)]{QuataertShiode2012} Quataert, E., \& Shiode, J.\ 2012, \mnras, 423, L92

\bibitem[Ryde \& Sch{\"o}ier(2001)]{RydeSchoier2001} Ryde, N. \& Sch{\"o}ier, F.~L.\ 2001, \apj, 547, 384

\bibitem[Saladino et al.(2018)]{Saladinoetal2018} Saladino, M.~I., Pols, O.~R., van der Helm, E., Pelupessy, I., \& Portegies Zwart S.\ 2018, \aap, 618, A50 

\bibitem[Shiode \& Quataert(2014)]{ShiodeQuataert2014} Shiode, J.~H., \& Quataert, E.\ 2014, \apj, 780, 96

\bibitem[Smith et al.(2010)]{Smithetal2010} Smith, N., Miller, A., Li, W., et al.\ 2010, \aj, 139, 1451

\bibitem[Soker(2008)]{Soker2008eff} Soker, N.\ 2008, \na, 13, 491

\bibitem[Soker \& Gilkis(2017)]{SokerGilkis2017} Soker, N. \& Gilkis, A.\ 2017, \mnras, 464, 3249

\bibitem[Strotjohann et al.(2021)]{Strotjohannetal2021} Strotjohann, N.~L., Ofek, E.~O., Gal-Yam, A., et al.\ 2021, arXiv:2010.11196 

\bibitem[Svirski \& Nakar(2014)]{SvirskiNakar2014} Svirski, G., \& Nakar, E.\ 2014, \apjl, 788, L14

\bibitem[Tartaglia et al.(2016)]{Tartagliaetal2016} Tartaglia, L., Pastorello, A., Sullivan, M., et al.\ 2016, \mnras, 459, 1039

\bibitem[van Marle et al.(2009)]{vanMarleetal2009} van Marle, A.~J., Owocki, S.~P., \& Shaviv, N.~J.\ 2009, \mnras, 394, 595

\bibitem[Vlemmings et al.(2002)]{Vlemmingsetal2002} Vlemmings, W.~H.~T., Diamond, P.~J., \& van Langevelde, H.~J.\ 2002, \aap, 394, 589

\bibitem[Wood \& Karovska(2006)]{WoodKarovska2006} Wood, B.~E. \& Karovska, M.\ 2006, \apj, 649, 410

\bibitem[Yaron et al.(2017)]{Yaronetal2017} Yaron, O., Perley, D.~A., Gal-Yam, A., et al.\ 2017, Nature Physics, 13, 510


\end{thebibliography}
\end{document}